\documentclass{svproc}

\usepackage{amsmath,amssymb}
\usepackage{graphicx}
\usepackage{booktabs}
\usepackage{multirow}
\usepackage{placeins}   
\usepackage{tikz}
\usetikzlibrary{arrows.meta, positioning, shapes.geometric}
\usepackage{capt-of} 
\usepackage{url}

\usepackage{hyperref} 
\renewcommand{\thetable}{\Roman{table}}

\title{ORViT-DR: Ordinality-Aware Hybrid ViT for Low-Resolution Diabetic Retinopathy Grading}
\subtitle{Accepted at ECCT 2026}
\titlerunning{Ordinality-Aware Hybrid ViT for Low-Resolution DR Grading}
\authorrunning{Kundu et al.}
\author{
Soumit Kumar Kundu\inst{1} \and
Nabil Ashab\inst{1} \and
Bidhan Biswas\inst{1} \and
Shahadat Hossain Sohag\inst{1} \and
Saif Mahmud Parvez\inst{1} \and
Souvik Kumar Kundu\inst{2} \and
Zunayed Ahmed Rafi\inst{3}
}

\institute{
Department of Computer Science and Engineering, 
Dhaka International University, Dhaka, Bangladesh
\email{soumitkumar4@gmail.com, nabilashab@gmail.com, bidhan.biswas970@gmail.com, shahadat.h.sohag8@gmail.com, saif.mahmud.parvez@gmail.com}
\and
KUET, Khulna, Bangladesh
\email{souvik3.cse@gmail.com}
\and
Bangladesh Bank, Dhaka, Bangladesh
\email{zunayed.ahmedrafi@gmail.com}
}

\usepackage[htt]{hyphenat}
\begin{document}
\mainmatter
\maketitle

\begin{abstract}
Diabetic retinopathy (DR) is one of the main causes of impaired vision. A good and reliable automated grading system can make the screening process safer and more accurate. Because DR stages progress gradually, the task of grading disease severity naturally follows an ordinal structure in which neighboring classes share similar visual characteristics. In this study, ORViT-DR, a hybrid deep learning framework, is designed to improve DR grading from low-resolution retinal images. The proposed approach combines convolutional feature extraction with transformer-based global context modeling through a pre-trained ViT-Hybrid backbone, which integrates BiT-ResNetv2 with a Vision Transformer architecture. The approach is tested on the RetinaMNIST subset of the MedMNISTv2 dataset, which contains 28×28 retinal fundus images annotated with five levels of disease severity. To promote stable training and better feature learning, the training strategy applies progressive layer unfreezing, layer-wise learning rate decay, exponential moving average (EMA) parameter updates, and ensemble-based prediction during inference. Experimental results on the official RetinaMNIST test set show that the proposed method achieves 57.00\% classification accuracy, along with a quadratic weighted kappa score of 0.5963 and a macro-F1 score of 0.4293. These results suggest that hybrid CNN–Transformer architectures can provide effective representations for DR grading with with competitive ordinal agreement measured by QWK.

\end{abstract}

\section{Introduction}

Diabetes mellitus is a critical health condition. It occurs from the body's inability to produce a proper amount of insulin to absorb glucose. Diabetic Retinopathy (DR) is the result of damaging blood-vessels of the retina due to diabetes. It causes low vision. If it is not noticed early, it can lead to vision impairment. For detection, ophthalmologists examine retinal fundus images to find lesions and other visual signs associated with DR. As screening initiatives can generate a huge number of images, these make evaluation lengthy. Screening them manually takes a lot of time. Automated grading systems can be useful here. They can help reduce the workload of doctors and help critical patients who may need better care.

In spite of recent progress, building reliable DR grading models is still not easy. Many important signs in retinal images can be missed as they are very subtle. Also, image quality is not always consistent—lighting, noise, and capture conditions often change them. It becomes more challenging when the dataset is class-imbalanced, small or low-resolution. The RetinaMNIST subset of MedMNISTv2 \cite{medmnist2023,medmnist2021} is a good example. The images are only 28×28 pixels, which have five severity levels with unevenly distributed classes.

Deep learning models\cite{resnet2018,densenet2019} plays a vital role in medical image classification. Convolutional neural networks (CNNs) \cite{lecun1998lenet,cnnadv2018} are widely used because they catch local spatial features from images effectively. More recently, Vision Transformers (ViTs) \cite{vit2021} have drawn significant interest because of their self-attention mechanisms. They can capture the global context of different regions of an image.  But, as they are transformer-based, they require large amounts of training data and often struggle with limited or low-resolution datasets. Another approach is using hybrid architectures. A combination of CNN feature extraction and transformer-based global reasoning\cite{medvit2023,hybridtransformer2024} introduces a new approach. ORVit-DR is introduced to study the feasibility of these types of hybid CNN-Tran\-former framework. It's a pretrained CNN-ViT backbone with local and global feature extraction, which could leverage the benefits from low-resolution rental images for grading diabetic retinopathy effectively. And for this investigation, RetinaMNIST could be very useful due to its limited training data and low-resolution images.

\section{Related Work}

\subsection{Benchmark Datasets for Medical Image Classification}

Benchmark Datasets e.g MedMNIST \cite{medmnist2023,medmnist2021} dataset collection plays a vital role in deep learning research. Images in this dataset is downscaled to 28×28 pixel resolution. It helps researchers with low computation power to use deep learning models. MedMNISTv2\cite{medmnist2023} is another dataset with both 2D and 3D images. It contains a group of datasets  such as retinal fundus images, blood cell microscopy, breast ultrasound, and histopathology images. It has become a great tool to compare models, AutoML systems, and transfer learning methods in biomedical image analysis\cite{texturebiomed2022}.

\subsection{Convolutional Neural Networks in Medical Imaging}

Convolutional Neural Networks (CNNs) and other deep learning models are now used a lot in medical imaging. The main reason is because they can find patterns from images in a layered way. They catch simple features first, then gradually move to the complex ones. This works great for medical image data. Pretrained models like ResNet, DenseNet, and other convolutional neural networks (CNNs) \cite{resnet2018,densenet2019,imagenet2015}have shown considerable efficacy in different types of medical image classification. CNNs excel at recognizing localized patterns and textures in images\cite{cnnadv2018,resnet2018}, which is beneficial for detecting lesions or structural anomalies. But over-dependence on local context often hampers CNN-based models' ability to capture global context.

\subsection{Vision Transformers in Computer Vision}

In the last few years, transformer-based models are being used in computer vision effectively. Vision Transformers (ViTs)\cite{vit2021,imagenet21k2021} convert input images as sequences of patches, uses self-attention for interaction among patches. This design help network's capacity to effectively understand the global context of an image. This feature is unique and helps to get overall insight in medical image classification.

\subsection{Transfer Learning with Pretrained ViT Models}

ViT models are computation heavy. So, they aren't usually trained from scratch. Pre-trained models on extensive datasets\cite{imagenet2015,imagenet21k2021} help in this case. This transfer learning strategy helps the model to use previously acquired knowledge in new architecture with minimal fine-tuning.

\subsection{Hybrid CNN–Transformer Architectures}

Instead of depending only on convolutional or transformer-based designs, hybrid models\cite{medvit2023,hybridtransformer2024} can be a great solution. These architectures frequently contain convolutional layers in the initial phases  to extract local features, transformer modules to fetch global contextual information. The idea here is fairly simple. Convolutional networks are good at learning structured, local features, while transformers are better at handling global relationships. By combining both, the model can benefit from each side. For this reason hybrid CNN–Transformer approaches are getting more attention in medical imaging tasks. This is especially true when datasets are small. It also helps when visual differences are very subtle. 

\subsection{Evaluation Metrics and Training Strategies in DR Analysis}

Another important thing in diabetic retinopathy (DR) analysis is that the disease stages follow an order. The severity does not jump randomly; it increases step by step. If a model predicts a nearby stage, the mistake is usually less serious than predicting a stage that is far off. So, accuracy alone is not always enough. Metrics like quadratic weighted kappa (QWK) are also used, since they consider how close the predicted label is to the actual one and helps to get real clinical grading. Another issue is with smaller and class-imbalanced datasets like RetinaMNIST. Training can become unstable, and models may not generalize well. To deal with this,  data augmentation, adjusting the learning rate over time, regularization, and using methods like exponential moving averages are used. All of these help make the model more stable and improve its overall performance.

Ordinal regression methods explicitly model ordered class relationships. Representative approaches include ranking-based formulations, cumulative probability methods, and distance-aware losses that penalize larger label gaps more strongly than adjacent errors. Such methods are well suited to medical grading problems where neighboring mistakes are clinically less severe than distant misclassifications. In the present work, weighted cross-entropy is employed for optimization and QWK is used as an ordinal-sensitive evaluation metric, while explicit ordinal objectives remain an important direction for future study.


\section{Research Motivation and Objectives}

Diabetes is an increasing concern in the world right now because of its ability to badly affect different organs. Early detection and treatment can be very effective against it. Among them, diabetic retinopathy is one of the most common and dangerous conditions because it can easily advance into blindness. So a diagnosis is very crucial for it, and an inaccurate assessment can lead to a catastrophic disaster. Deep learning has already proven super accurate and robust performance in some of the sectors of image classification. But their performance can be very limited in medical image diagnosis because of a limited dataset and low, noisy captured images. Another reason is that publicly available datasets are heavily imbalanced and often represent bias towards specific portions. Effective classification despite these limitations is the key to medical image diagnosis. That's why RetinaMNIST is fit for this because it shows all the limitations described above, like low resolution image (28x28), an uneven dataset, etc.

Motivated by these challenges, this work proposes ORViT-DR, a hybrid CNN–Transformer framework designed to improve DR severity grading under constrained data conditions. This study investigates whether combining convolutional feature extraction with transformer-based global context modeling can improve classification accuracy, without needing extensive retraining on large datasets.

In particular, the work is guided by the following research questions:

\begin{itemize}
    \item Can a hybrid CNN–Transformer backbone outperform conventional CNN and AutoML baselines on RetinaMNIST without requiring large-scale retraining from scratch?
    \item Which fine-tuning and stabilization choices (progressive unfreezing, layer-wise learning rate decay, EMA, MixUp/CutMix) contribute most to stable learning in low-data regimes?
    \item Does ensemble-based inference improve ordinal agreement (QWK) in addition to overall classification accuracy?
\end{itemize}


\section{Datasets Utilized in This Research}

The dataset used in this research is RetinaMNIST from MedMNISTv2 via the Hugging Face dataset identifier albertvillanova/MedMNISTv2 with configuration RetinaMNIST. The dataset comprises \textbf{1,600} fundus images(split mentioned in \textbf{Table~\ref{tab:table_1}}) labeled into five ordinal DR grades. 

\begin{table}[htbp]
\centering
\caption{Baseline dataset statistics}
\label{tab:table_1}
\begin{tabular}{lccc}
\toprule
\textbf{Split} & \textbf{Samples} & \textbf{Classes} & \textbf{Image Resolution} \\
\midrule
Train & 1080 & \multirow{3}{*}{\begin{tabular}{c}
5 (No DR, Mild, Moderate,\\
Severe, Proliferative)
\end{tabular}} 
& \multirow{3}{*}{\begin{tabular}{c}
28$\times$28 (resized to\\
384$\times$384 by the\\
model processor)
\end{tabular}} \\
\cmidrule(lr){1-2}
Validation & 120 &  &  \\
\cmidrule(lr){1-2}
Test & 400 &  &  \\
\bottomrule
\end{tabular}
\end{table}

\begin{figure}[ht]
\centering
\includegraphics[width=0.8\linewidth]{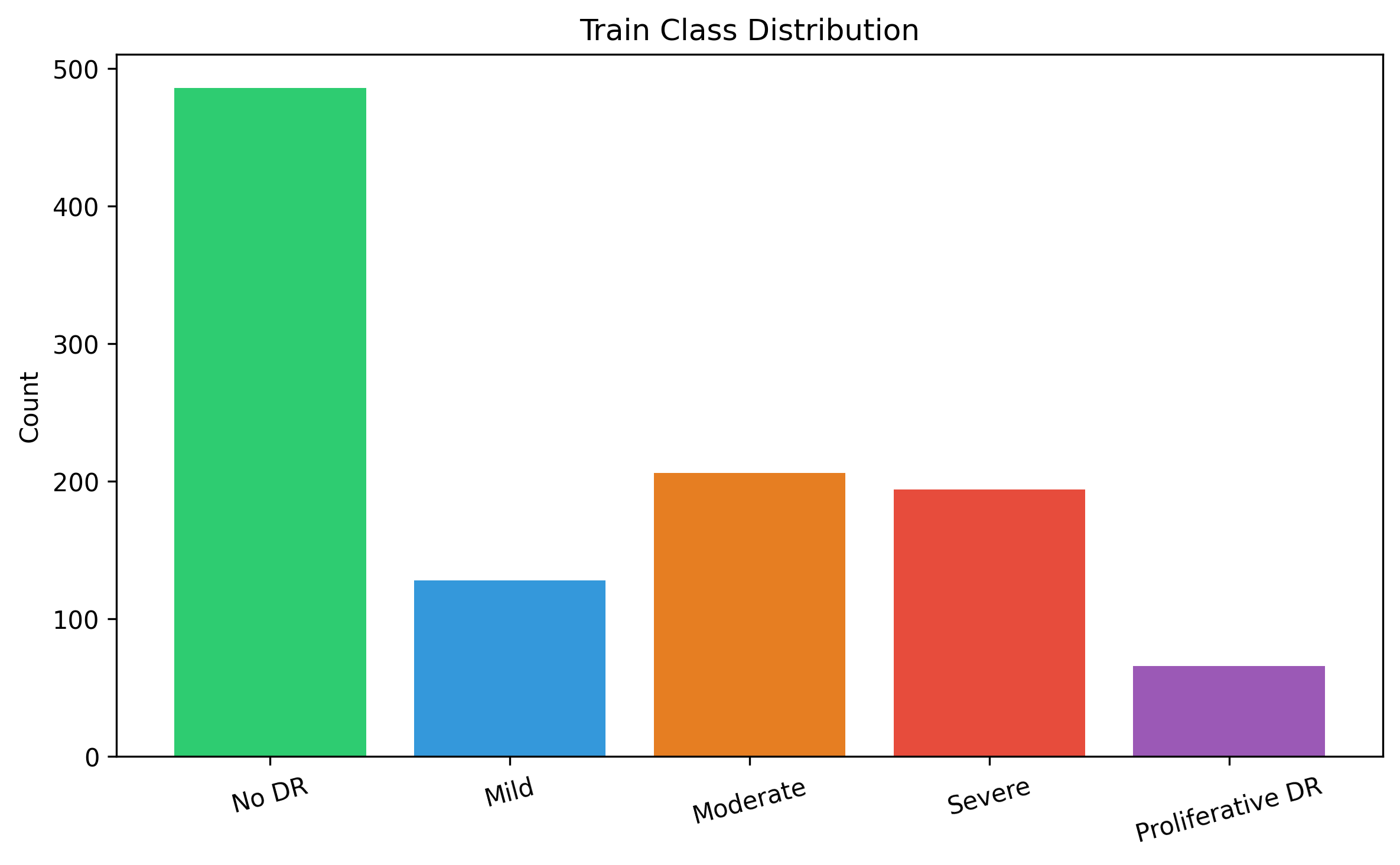}
\caption{Class distribution of the RetinaMNIST training split (imbalanced toward No DR).}
\label{fig:image_1}
\end{figure}

Different training class distribution of RetinaMNIST is shown in \textbf{Fig.~\ref{fig:image_1}}. No DR dominates other classes significantly.

\section{Methodology and Implementation}

\subsection{Data Pre-processing}

Although RetinaMNIST images are 28×28, the hybrid backbone vit-hybrid-base-bit-384 expects 384×384 inputs. Images are resized using the model processor for input compatibility and transfer-learning alignment. This resizing step heps effective adaptation of the pretrained backbone to the target benchmark. The model’s image processor is used to resize and normalize each image. Augmentations (RandAugment, color jitter, affine transforms, flips) are applied before processor resizing. To improve robustness, MixUp and CutMix are used during fine-tuning \cite{mixup2018,cutmix2019}.

Weighted cross-entropy (with smoothing). With class weights $w_c$, one-hot labels $y_c$, 
and predicted probabilities $p_c$, the objective is:

\begin{equation}
\mathcal{L}_{CE} = -\sum_{c=1}^{C} w_c \, y_c \log(p_c)
\end{equation}

Label smoothing replaces hard one-hot targets with a convex combination of the one-hot label and a uniform prior.

\subsection{Proposed Architecture}

The ORViT-DR architecture, as proposed, employs a hybrid CNN–Transformer pipe-line, which is structured to extract both local spatial characteristics and global contextual relationships inherent in retinal fundus images. This comprehensive framework is organized into five principal stages: input processing, pre-processing and augmentation, feature encoding via dual encoders, classification, and inference stabilization.

\begin{figure}[ht]
\centering
\includegraphics[width=1\linewidth]{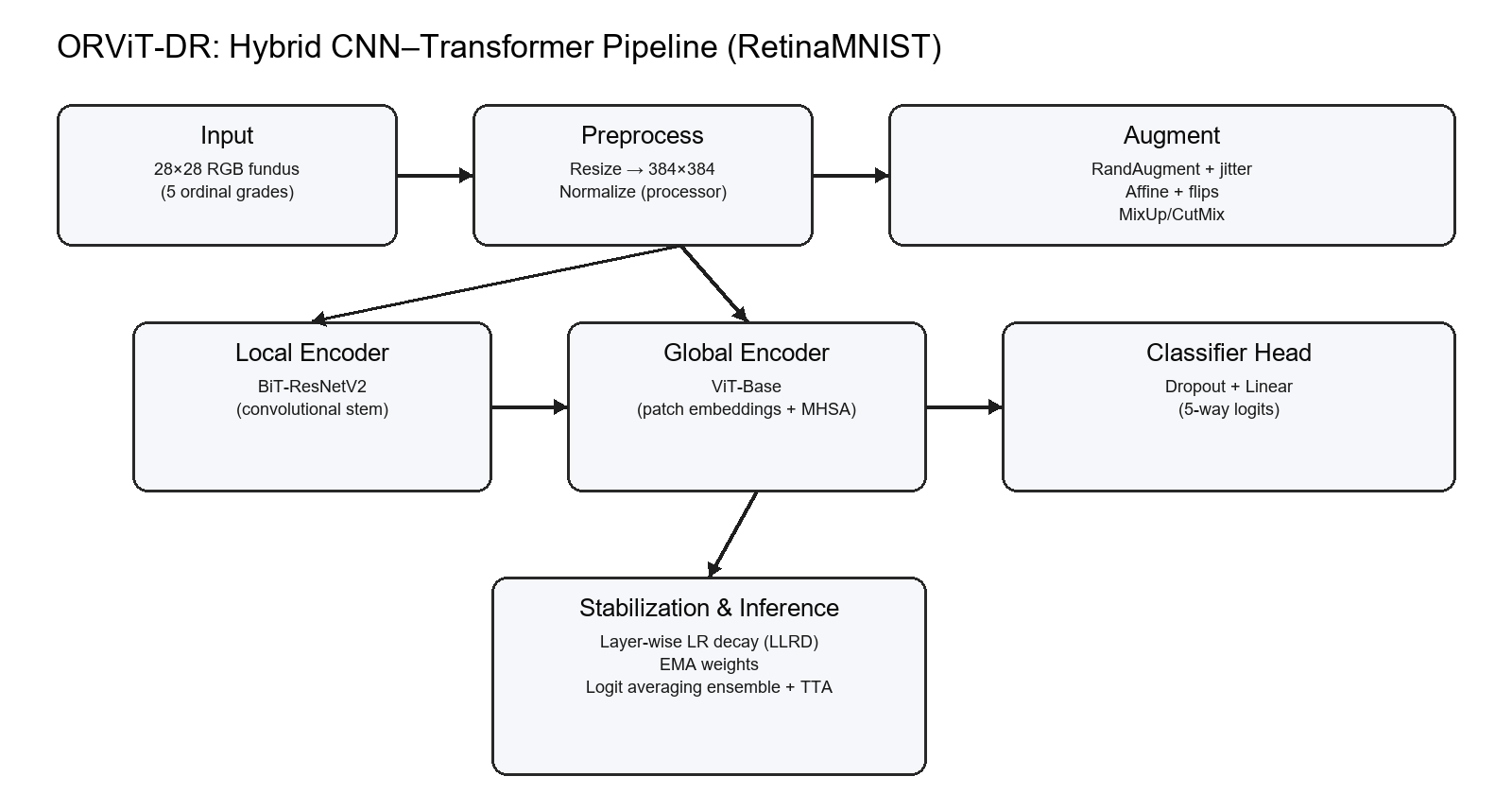}
\caption{ORViT-DR whole architecture. It is based on a hybrid CNN–Transformer backbone.}
\label{fig:image_2}
\end{figure}

The process begins with retinal fundus images from the RetinaMNIST dataset. Each image is a 28 × 28 RGB representation, corresponding to one of five ordinal diabetic retinopathy grades. Because the selected pretrained backbone operates on 384$\times$384 inputs, images are resized to match that. This step is used for architectural compatibility rather than to generate additional visual detail. Normalization follows with previously fixed parameters after image resizing. Stability of the training process is ensured with this distribution. The whole architecture is shown in \textbf{Fig.~\ref{fig:image_2}}.

\subsection{Core Settings}

Core settings of the ORViT-DR architecture.
\begin{itemize}
\item \textbf{Task:} 5-class diabetic retinopathy grading using RetinaMNIST.
\item \textbf{Data split:} train/val/test = 1080/120/400
\item \textbf{Input:} Images are originally 28×28, but resized to 384×384 for the model.
\item \textbf{Model:} Hybrid ViT (vit-hybrid-base-bit-384): BiT-ResNetV2 feature extractor, ViT-base transformer encoder
\item \textbf{Learning rate:} Around 5e-4 in the first stage, then smaller (3e-5 backbone, 3e-4 head) with decay across layers.
\item \textbf{Loss \& metrics:} Weighted cross-entropy loss. Evaluation uses accuracy, macro-F1, and QWK.
\end{itemize}

\subsection{Training}
The training procedure is divided into two phases:
\begin{itemize}

\item \textbf{Head-only training:} Initially, only the classification head is trained while the backbone remains frozen.

\item \textbf{Backbone training:} Subsequently, part of the backbone is unfrozen and training continues. A batch size of 8 is used in this phase, along with a cosine learning rate schedule.
\end{itemize}

\subsection{Regularization}
Label smoothing is done to prevent the model from becoming overconfident. Also dropout is applied to keep the model from relying too much of specific features, and gradient clipping is applied to keep the ongoing training stable.

\subsection{Augmentation}
To reduce overfitting and maximize generalization due to a heavy dataset.
\begin{itemize}
\item \textbf{RandAugment:\cite{cubuk2020randaugment}} It is used to induce alterations to the images. Slightly changing color or intensity randomly helps the model generalize better.
\item \textbf{Color jitter:} Adding a slight variation of brightness, contrast, and color of the images helps to handle different visual conditions.
\end{itemize}

\subsection{Fine Tuning}
Different fine-tuning approaches have been taken after analyzing the training progress and the progression of epochs. Continuous monitoring and observation have been made to check if any overfitting is occurring or not.
\begin{itemize}
\item \textbf{Cutmix:}\cite{cutmix2019} In this approach, a small patch from one image is taken and pasted onto another image. The label is also adjusted. It encourages the model to pay attention to different regions instead of depending on just one part.
\item \textbf{Mixup:\cite{mixup2018}} Two images as well as their labels are blended together. This makes the model become smoother and less noise-prone.
\end{itemize}

\subsection{Inference}

During inference, predictions from several high-performing checkpoints are averaged to improve stability. In addition, test-time augmentation (TTA) is applied by generating slightly modified versions of the same input image and averaging their predictions. This combination helps reduce variance in the final outputs and leads to more consistent predictions. The hybrid design of ORViT-DR, which integrates both local and global feature representations, further supports robust decision-making on low-resolution retinal images.



\subsubsection{vit-hybrid-base-bit-384}

Here \textit{google/vit-hybrid-base-bit-384} model has been used as hybrid setup. BiT-style ResNetV2  is used to extract local textures, features, tiny lesions. Then ViT-base transformer encoder is to used for global context. The ViT-base linked information from different regions. Thus, the model gets a mix of both local and global feature extraction which especially help for low-resolution images.

\subsubsection{Key Components of the Hybrid ViT Architecture}
The architecture uses a convolutional stem, such as BiT or ResNetV2 \cite{kolesnikov2020bit,he2016resnetv2}, to extract local features and improve translation robustness\cite{medrdf2022}.
\begin{itemize}
\item Transformer encoder for long-range dependency modeling across the image grid.

\item Lightweight classification head (Dropout + Linear) to mitigate overfitting in low-data regimes.

\end{itemize}

\subsection{Fine-Tuning Strategy}

The proposed ORViT-DR framework is trained using a two-stage fine-tuning strategy designed to stabilize optimization while adapting the pretrained backbone to the target retinal dataset.Layer-wise learning rate decay (LLRD) is applied so earlier pretrained layers receive smaller updates while later layers adapt more rapidly. Exponential moving average (EMA) with decay 0.999 is maintained during training to smooth parameter updates and improve generalization. A practical consideration for implementation is using full precision (FP32) training, because mixed precision can lead to numerical instability. The GroupNorm layers in the BiT backbone are more likely to overflow when using half precision. This can cause NaN values during training. So, to make sure that convergence is dependable, FP32 is used for all testing. Training configuration and fine-tuning strategy is given in \textbf{Table~\ref{tab:table_2}}

\begin{table}[htbp]
\centering
\caption{Training configuration and fine-tuning strategy}
\label{tab:table_2}
\begin{tabular}{lp{10cm}}
\toprule
\textbf{Component} & \textbf{Setting (from implementation)} \\
\midrule
Stage 1 & Head-only training (warmup + cosine), learning rate $5\times10^{-4}$, batch size 8 \\

Stage 2 & Partial unfreezing of transformer encoder + LayerNorm + head \\

LLRD & Decay factor 0.85, base LR $3\times10^{-5}$, head LR $3\times10^{-4}$ \\

Regularization & Label smoothing 0.1, dropout in head, gradient clipping (1.0) \\

MixUp/CutMix & MixUp $\alpha = 0.3$; CutMix $\alpha = 1.0$, randomly sampled per batch \\

EMA & Decay 0.999 (applied on trainable parameters) \\

Ensemble & Uniform logit averaging over top checkpoints; optional TTA views \\
\bottomrule
\end{tabular}
\end{table}

In the first stage, only the classification head is trained while keeping the backbone networks frozen. This head-only training phase acts as a warm-up step that allows the classifier to adapt to the target task without disrupting the pretrained representations. During this stage, a cosine learning rate schedule is applied with an initial learning rate of $(5 \times 10^{-4})$, and training is performed with a batch size of 8.

In the subsequent phase, partial unfreezing is implemented to facilitate a more profound adaptation of the model. More precisely, the transformer encoder and its corresponding LayerNorm layers are unfrozen, alongside the classification head. This prevents the model from not losing what it already learned from pretraining and adapting that knowledge for DR grading.

For stable training, layer-wise learning rate decay (LLRD) is introduced. This means the earlier layers change more slowly, while the later layers are allowed to adjust more. A decay factor of 0.85 is used and a learning rate of $(3 \times 10^{-5})$ for the backbone and $(3 \times 10^{-4})$ for the classification head. Label smoothing (0.1) slightly softens the target labels. Dropout is added to the classification head, and gradient clipping (with a max norm of 1.0) is used to avoid unstable updates. These helps reduce overfitting. For data augmentation, both MixUp and CutMix are used. MixUp ($\alpha = 0.3$) blends two images together, while CutMix ($\alpha = 1.0$) swaps patches between images. Both are not used at once, only one is chosen randomly for each batch, which helps the model learn more general patterns. EMA with decay 0.999 are maintained during training. This keeps the parameter updates smoother and keeps slightly more consistent performance. This leads to increased stability and often results shows better generalization performance during evaluation.

\subsection{Ablation Protocol}
To assess the contribution of key components, controlled ablations for layer-wise learning rate decay (LLRD), EMA, and checkpoint ensembling were done. For the LLRD control, seed and initialization were fixed and  Phase-2 fine-tuning was reran with LLRD disabled while keeping all other settings unchanged. This allowed isolating the effect of each component without confounding factors.

\section{Evaluation Approach}

Evaluation is done using three complementary metrics:

\begin{itemize}
\item Accuracy to match MedMNIST benchmark reporting.
\item Macro-F1 \cite{f1score2020} to account for class imbalance.
\item QWK to measure ordinal agreement and penalize distant misclassifications.
\end{itemize}

For ROC analysis, one-vs-rest ROC curves is computed \cite{roc1997} per class using the ensemble softmax probabilities and report per-class AUC.

To assess training stability, additional experiments were conducted across three independent random seeds. Moderate variability was observed, which is expected given the limited size and low resolution of the dataset. While individual runs vary, the best-performing configuration achieves the peak performance reported in the main results. A more comprehensive multi-seed evaluation and statistical analysis is considered for future work.

\section{Result Analysis}

\subsection{Confusion Matrix}

\begin{figure}
\centering
\includegraphics[width=0.8\linewidth]{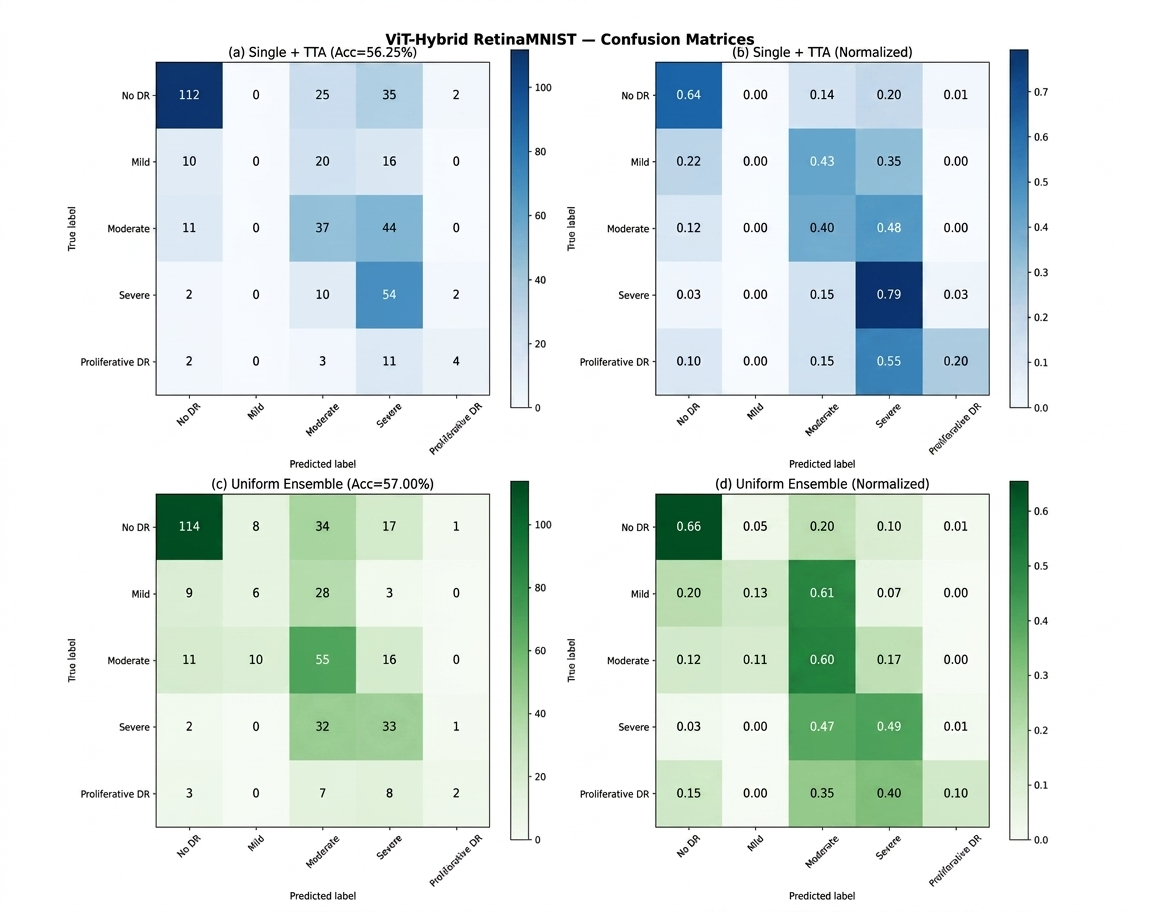}
\caption{Confusion matrices for the best single checkpoint (56.25\%) and top-3 uniform ensemble (57.00\%).}
\label{fig:image_3}
\end{figure}

The confusion matrices in \textbf{Fig.~\ref{fig:image_3}}.  indicate that most errors occur between adjacent grades (e.g., Mild $\leftrightarrow$ Moderate, Moderate $\leftrightarrow$ Severe), consistent with the ordinal nature of DR. The ensemble improves stability in moderate and severe predictions but remains challenged on the minority proliferative class.

\begin{figure}[ht]
\centering
\includegraphics[width=1\linewidth]{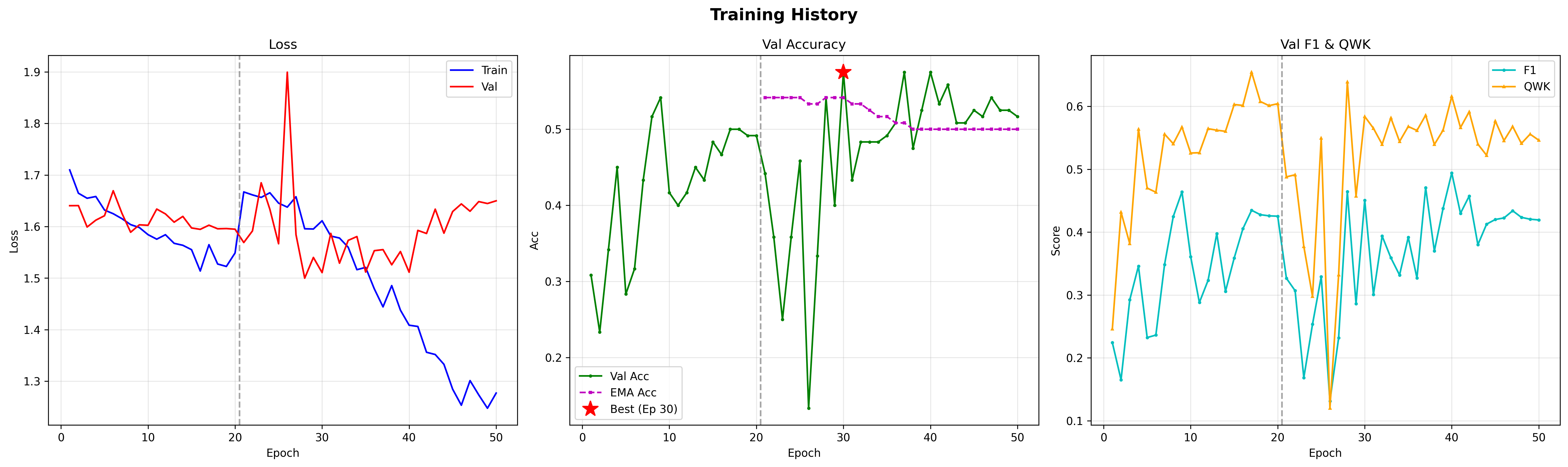}
\caption{Training curves produced by the implementation: (a) validation accuracy and (b) train/validation loss across stages.}
\label{fig:image_4}
\end{figure}

\subsection{ROC Curve}

\begin{figure}[ht]
\centering
\includegraphics[width=0.8\linewidth]{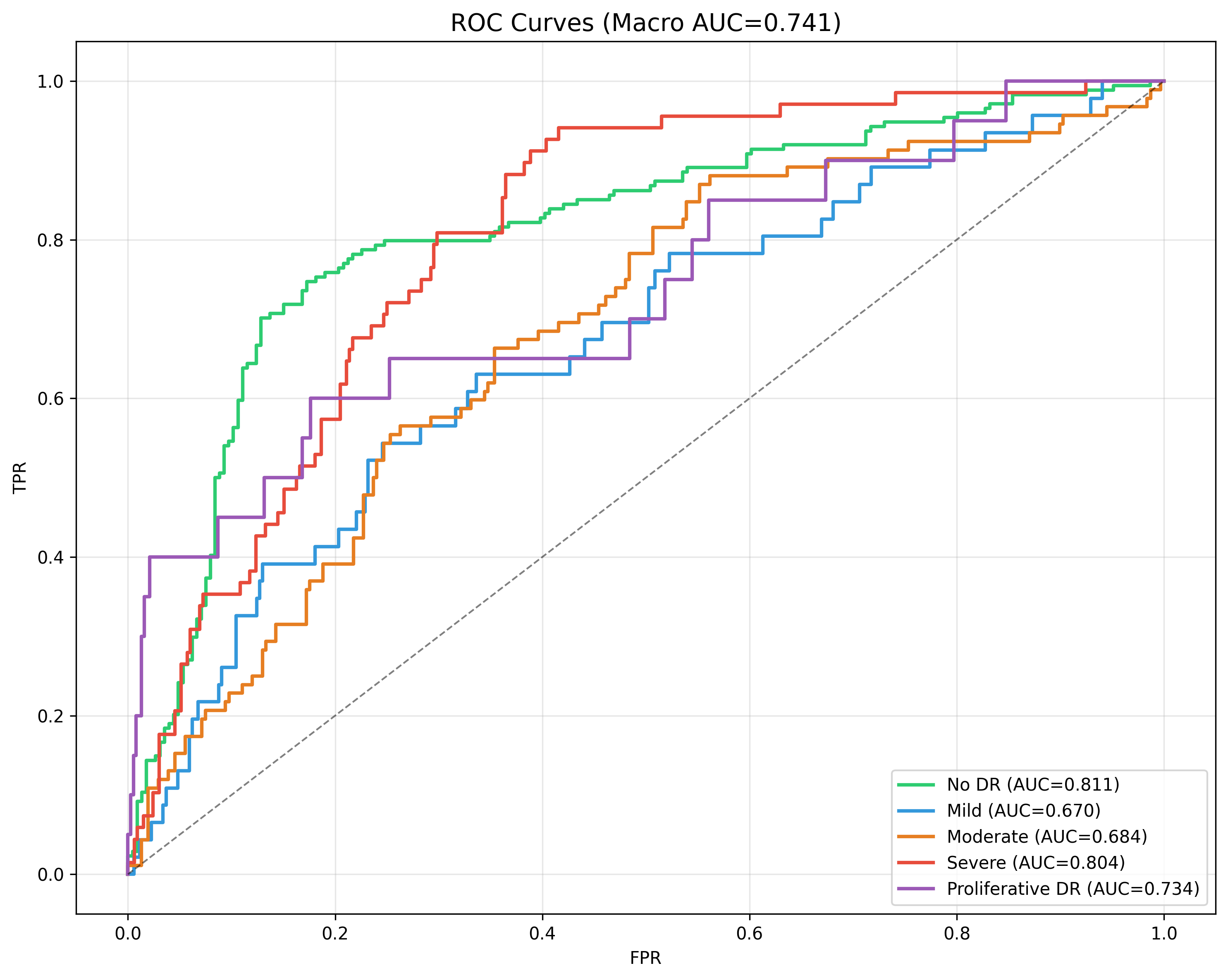}
\caption{One-vs-rest ROC curves for the top-3 ensemble. The per-class AUCs are shown in the legend.}
\label{fig:image_5}
\end{figure}

The top-3 ensemble yields strong separability for No-DR and Severe classes (AUC $\approx 0.83$), while Mild remains harder (AUC $\approx 0.68$), likely due to subtle visual cues and class imbalance. This is mentioned in ROC curve \textbf{Fig.~\ref{fig:image_5}}.

\subsection{Ablation Study}

Removing LLRD led to a substantial performance drop (--8 pp for single-model accuracy), confirming its importance for stable fine-tuning. Checkpoint ensembling consistently improved performance, while EMA contributed to training stability. These findings validate the role of key components in the proposed framework.

\begin{table}[t]
\centering
\caption{Ablation study under a fixed seed setting (seed 42). Results are reported as relative performance changes.}
\label{tab:ablation}
\begin{tabular}{lcc}
\toprule
Configuration & $\Delta$ Single + TTA (pp) & $\Delta$ Ensemble + TTA (pp) \\
\midrule
No LLRD vs. With LLRD & -8.00 & -2.25 \\
\bottomrule
\end{tabular}
\end{table}

\section{Grad-CAM Analysis}

To improve model interpretability, Gradient-weighted is applied Class Activation Mapping (Grad-CAM) to visualize image regions that most strongly influence class decisions. Grad-CAM produces a class-specific saliency map by combining feature activations from a target convolutional layer with gradient-based importance weights.

\begin{figure}[ht]
    \centering
    \includegraphics[width=0.98\linewidth]{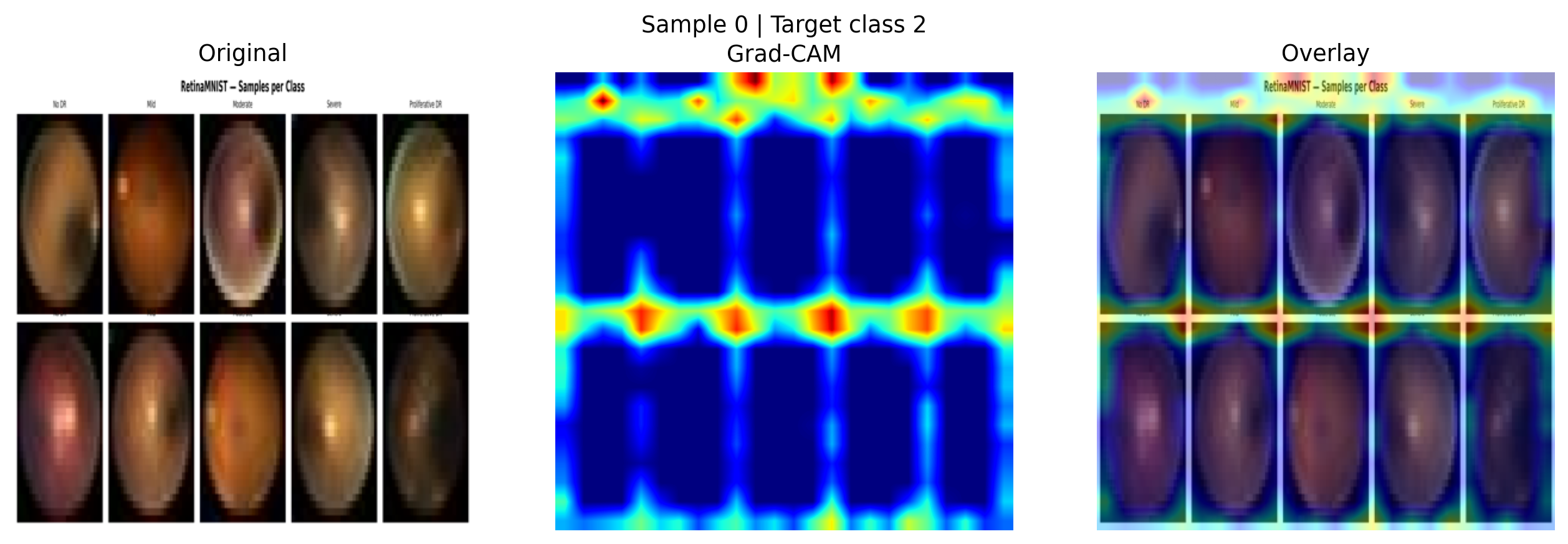}
    \caption{Representative Grad-CAM visualizations on RetinaMNIST samples. From left to right: original image, Grad-CAM heatmap, and heatmap overlay. Warmer regions indicate a stronger positive contribution to the predicted class.}
    \label{fig:gradcam_examples}
\end{figure}

In this work, Grad-CAM is generated from a deep convolutional layer in the hybrid retina classifier, and heatmaps are upsampled and overlaid on the corresponding fundus images. These visualizations is then used to verify that high-response regions align with clinically relevant retinal structures and lesion patterns. Figure~\ref{fig:gradcam_examples} shows representative examples of original images, Grad-CAM maps, and overlays.

\section{Comparison with Benchmark Approaches}

\textbf{Table~\ref{tab:table_3}} compares ORViT-DR against standard CNN and AutoML baselines reported for RetinaMNIST in the MedMNIST benchmark literature.

Compared to CNN/AutoML\cite{automlsurvey2021,autokeras2019} baselines ($\approx$50--53\% accuracy), ORViT-DR improves test accuracy by up to ~4 pp and improves ordinal agreement as reflected by QWK. The hybrid backbone achieves transformer-level performance while benefiting from CNN inductive biases.

\begin{table}
\centering
\caption{Benchmark comparison on RetinaMNIST}
\label{tab:table_3}
\begin{tabular}{p{4cm}cp{5.5cm}}
\toprule
\textbf{Method} & \textbf{RetinaMNIST Acc (\%)} & \textbf{Notes} \\
\midrule
ResNet-18 (28) & 52.4 & Baseline from MedMNIST \\

ResNet-18 (224) & 49.3 & Baseline with 224-resized input \\

ResNet-50 (28) & 52.8 & Baseline from MedMNIST \\

ResNet-50 (224) & 51.1 & Baseline with 224-resized input \\

auto-sklearn & 51.5 & Classical AutoML baseline \\

AutoKeras & 50.3 & Neural search AutoML baseline \\

Google AutoML Vision & 53.1 & Strong AutoML baseline \\

\textbf{ORViT-DR (best single checkpoint)} & 56.25 & Hybrid ViT backbone; Dropout head; QWK=0.6052 \\

\textbf{ORViT-DR (top-3 uniform ensemble)} & 57.0 & Logit averaging; Macro-F1=0.4293; QWK=0.5963 \\

\bottomrule
\end{tabular}
\end{table}

\section{Conclusion \& Future Works}

This manuscript presents ORViT-DR, a hybrid CNN–Transformer approach for low-resolution DR grading on RetinaMNIST. The method combines a pre-trained hybrid backbone with progressive unfreezing, layer-wise learning-rate decay, EMA stabilization, and ensemble inference. ORViT-DR achieves 57.00\% test accuracy, macro-F1\cite{f1score2020} 0.4293, and QWK 0.5963, outperforming conventional CNN and AutoML baselines reported for RetinaMNIST. 

Future work will focus on (i) improved imbalance handling (ordinal loss, class-balanced sampling) (ii) more comprehensive evaluation through repeated runs and statistical analysis (iii) additional comparisons with transformer-only and other modern backbones will also be explored.






\FloatBarrier

\bibliographystyle{splncs03_unsrt}
\bibliography{references_svproc_fixed}

\end{document}